\documentclass[%
preprint,
showpacs,
nobibnotes,
 amsmath,amssymb,
 aps,pra,
]{revtex4-1}

\usepackage{graphicx}% Include figure files
\usepackage{dcolumn}% Align table columns on decimal point
\usepackage{bm}% bold math
\newcommand{\vB}{\mbox{\boldmath $ B $}}

\newcommand{\vF}{\mbox{\boldmath $ F $}}
\newcommand{\vJ}{\mbox{\boldmath $ J $}}

\newcommand{\rr}{\mbox{\boldmath $ r $}}
\newcommand{\vU}{\mbox{\boldmath $ U $}}

\newcommand{\vvy}{\mbox{\boldmath $\hat y$}}
\newcommand{\vvz}{\mbox{\boldmath $\hat z$}}

\newcommand{\btimes}{\mbox{\boldmath $ \times $}}
\newcommand{\bnabla}{\mbox{\boldmath $ \nabla $}}

\begin{document}

\preprint{Submitted to Phys. Rev. E}

\title{Kelvin-Helmholtz versus Hall Magneto-shear instability in astrophysical flows}% Force line breaks with \\
%\thanks{}%

\author{Daniel O. G\'omez}\email[]{dgomez@df.uba.ar}
%\affiliation{Instituto de Astronom\'\i a y F\'\i sica del Espacio, C.C.~67 Suc.~28, 1428 Buenos Aires, Argentina.}
\altaffiliation[Also at ]{Departamento de F\'\i sica, Facultad de Ciencias Exactas y Naturales, Universidad de Buenos Aires, 1428 Buenos Aires, Argentina.}
%\homepage[]{http://astro.df.uba.ar}
%\thanks{}
%
\author{Cecilia Bejarano}
\affiliation{Instituto de Astronom\'\i a y F\'\i sica del Espacio, C.C.~67 Suc.~28, 1428 Buenos Aires, Argentina.}
%\thanks{}
\author{Pablo D. Mininni}
\affiliation{Departamento de F\'\i sica, Facultad de Ciencias Exactas y Naturales, Universidad de Buenos Aires \& Instituto de F\'\i sica de Buenos Aires, Ciudad Universitaria, 1428 Buenos Aires, Argentina.}

%\thanks{}
%

\date{\today}% It is always \today, today,

\begin{abstract}
We study the stability of shear flows in a fully ionized plasma. Kelvin-Helmholtz is a well known, macroscopic and ideal shear-driven instability. In sufficiently low density plasmas, also the microscopic Hall magneto-shear instability can take place. We performed three-dimensional simulations of the Hall-MHD equations where these two instabilities are present, and carried out a comparative study. We find that when the shear flow is so intense that its vorticity surpasses the ion-cyclotron frequency of the plasma, the Hall magneto-shear instability is not only non-negligible, but it actually displays growth rates larger than those of the Kelvin-Helmholtz instability.
\end{abstract}
%\pacs{95.30.Qd Magnetohydrodynamics and plasmas, 47.27.-i Turbulent flows, 52.35.Py Macroinstabilities}

\maketitle

\section{\label{sec:intro}Introduction}

Even though the large-scale behavior of most astrophysical plasmas is well described using magnetohydrodynamics (MHD), at sufficiently smaller scales non-fluidistic effects might become relevant. For instance, when in a fully ionized hydrogen plasma we reach scales as small as the ion skin depth $c/\omega_{pi}$ ($c$ being the speed of light and $\omega_{pi}$ the ion plasma frequency), the Hall effect becomes non-negligible. This is often the case in various dynamical processes taking place in low-density plasmas, such as in the interstellar medium. Since astrophysical flows are also characterized by very large Reynolds numbers, this in turn implies that a wide range of spatial scales are relevant to properly describe their dynamical behavior all the way from the macroscopic size of the problem to 
intermediate scales such as $c/\omega_{pi}$ and down to scales small enough where energy eventually dissipates. The role played by the Hall effect in a variety of astrophysical flows, has been studied extensively in the literature. The role of the Hall current in turbulent regimes \cite{mininni_2007}, its relevance in the generation of magnetic fields by dynamo action \cite{mininni_2005,mininni_2002}, or its importance in magnetic reconnection \cite{dmitruk_2012,dmitruk_2013}, are only a few examples. 

At sufficiently small scales, the large-scale dynamics is usually perceived as a macroscopic velocity gradient, and it is often modeled through a shear flow. The existence of shear flows is ubiquitous in astrophysics. It is of interest in a variety of problems such as
astrophysical jets propagating in the interstellar medium \cite{bodo_1994}, zonal flows being formed in the atmospheres of rotating planets like Jupiter \cite{hasegawa_1985} or in the interaction of solar CMEs with the interplanetary medium \cite{foullonetal2013}. The stability of shear flows has been extensively studied and reviewed in the pioneering work of Chandrasekhar (see  Ref.~[\onlinecite{chandra1961}]). It has also been studied in a variety of astrophysical problems, such as jet collimation \cite{begelman1984}, the dynamics of spiral arms in galaxies \cite{dwarkadas1996}, accretion disks \cite{balbus1998}, or the solar wind \cite{poedts1998}.

The paradigmatic instability in shear flows is the well-known Kelvin-Helmholtz instability (KHI). It is an ideal hydrodynamic instability, that converts the energy of the large-scale velocity gradients into kinetic and/or magnetic energy at much smaller scales, eventually driving a turbulent regime at these scales. The presence of a magnetic field parallel to the shear flow has a stabilizing effect, and can even stall the instability if the shear velocity jump is smaller than twice the Alfven speed \cite{lau1980}. On the other hand, an external magnetic field perpendicular to the shear flow has no effect on the linear regime of the instability, and it is simply advected by the flow. The Kelvin-Helmholtz instability plays an important role in several space physics and astrophysics problems, such as the interface between the solar wind and magnetospheres \cite{parker1958}, coronal mass ejections \cite{foullonetal2013}, the stability of jet propagation \cite{birkinshaw1997} or cometary tails \cite{brandt1979}. A general stability analysis in the presence of a magnetic field was carried out in Ref.~[\onlinecite{miura1982}] (see also Refs.~[\onlinecite{chandra1961}] and [\onlinecite{lau1980}]).

A relatively less known instability is the so-called {\it Hall magneto-shear instability} (Hall-MSI), which arises in plasmas embedded both in a shear flow and an external magnetic field perpendicular to the flow \cite{bejarano2011}. It is an ideal and microscopic instability, since it takes place at all wavelengths smaller than the ion skin-depth. A linear study of Hall-MSI for weakly ionized plasmas has also been reported \cite{kunz2008}, which also includes the role of ambipolar diffusion. Hall-MSI arises only when the shear flow vorticity is anti-parallel to the external magnetic field, and corresponds to the destabilization of the ion-cyclotron wave mode. In other words, it arises whenever the shear is steep enough to be larger than the ion-cyclotron frequency \cite{bejarano2011}, and therefore the free energy from the shear flow is invested in accelerating ions in their cyclotron motion. This instability might also play a role at the interface between astrophysical jets and their surrounding environment, just as it is also the case for KHI. Therefore, our goal in this paper is to setup a numerical experiment to allow these two instabilities (i.e., Hall-MSI and KHI) to compete. 

This paper is organized as follows. In Section~\ref{sec:hmhd} we present the so-called Hall-MHD equations, which are an extension of the traditional one-fluid MHD equations that includes the effect of the Hall current. In Section~\ref{sec:shear} we show these same equations in the case where the plasma is embedded both in an external large-scale shear flow and in a uniform magnetic field perpendicular to the flow. The shear flow is maintained by an external force that reaches an equilibrium with the viscous force. This exact equilibrium of the Hall-MHD equations is perturbed and its linear stability is studied in Section~\ref{sec:linear}. Two competing instabilities are obtained: the macroscopic Kelvin-Helmholtz instability is studied in Section~\ref{sec:kh}, while the microscopic Hall-MSI instability is addressed in Section~\ref{sec:hmsi}. A comparative study between the corresponding growth rates of these two instabilities is performed in Section~\ref{sec:discu}. Finally, the conclusions of the present work are listed in Section~\ref{sec:conclu}.

\section{\label{sec:hmhd}Hall-MHD equations}

The incompressible Hall-MHD equations for a fully ionized hydrogen plasma are the modified induction equation (i.e., with the addition of the Hall current) and the equation of motion (the Navier-Stokes equation),
\begin{eqnarray}
\frac{\partial\vB}{\partial t} & = & \bnabla\btimes\left[\left( \vU - \epsilon v_A\bnabla\btimes\vB\right)\btimes\vB\right] + \eta \nabla^2 \vB \label{eq:HallMHD}\\
\frac{\partial \vU}{\partial t} & = & - \left( \vU \cdot \bnabla \right) \vU + v_A^2 \left( \bnabla\btimes\vB\right) \btimes \vB - \bnabla P  + \nu \nabla^2 \vU + \vF . 
\label{eq:NS}
\end{eqnarray}
The velocity $\vU$ is expressed in units of a characteristic speed $U_0$, the magnetic field $\vB$ is in units of $ B_0$, and we also assume a characteristic length scale $L_0$ and a spatially uniform particle density $n_0$. The assumption of incompressibility is valid provided that the plasma velocity associated with the instabilities being considered, remains significantly smaller than both the Alfv\'en velocity and the speed of sound. Because of quasi-neutrality, 
the electron and the proton particle densities are equal, i.e., $n_e = n_i = n_0$. The (dimensionless) Alfven speed is then $v_A = B_0/\sqrt{4\pi m_i n_0}U_0$, while $\eta $ and $\nu$ are respectively the dimensionless magnetic diffusivity and kinematic viscosity. The parameter $\epsilon $ is the dimensionless ion skin depth, and measures the relative strength of the Hall effect, 
\begin{equation}
\epsilon = \frac{c}{\omega_{pi} L_0},
\label{eq:eps}
\end{equation}
where  $w_{pi}=\sqrt{4\pi e^2 n_0/m_i}$ is the ion plasma frequency.

These equations are complemented by the solenoidal conditions for both vector fields, i.e.,
\begin{equation}
\bnabla\cdot\vB = 0 = \bnabla\cdot\vU\ .
\label{eq:div}
\end{equation}

From a theoretical point of view, Hall-MHD corresponds to a two-fluid description of a fully ionized plasma: a positively charged ion species of mass $m_i$ moving with the velocity field $\vU (\rr , t) $, and negatively charged massless electrons with the velocity
\begin{equation}
\vU_e = \vU - \epsilon v_A\bnabla\btimes\vB ,
\label{eq:ue}
\end{equation}
which stems from Amp\`ere's Law (i.e., $\vJ =\frac{c}{4\pi} \bnabla\btimes\vB$) and from the expression for the electric current density for this two-fluid plasma: $\vJ = e n_0 (\vU - \vU_e)$. Note that it is a simplified version of a two-fluid description, since we are 
neglecting the mass of electrons. For this reason, the smallest scales covered by this description have to remain much larger than 
the scale of electron inertia $c/\omega_{pe}$ ($\omega_{pe}=\sqrt{4\pi e^2n_0/m_e}$ is the electron plasma frequency), which is 
determined by the electron mass $m_e$.

\section{\label{sec:shear}Shear-driven Hall-MHD equations}
Let us assume that the plasma is subjected to an externally applied shear flow given by
\begin{equation}
\vU_0 = u_0(x) \vvy ,
 \label{eq:shear}
\end{equation}
so that the total velocity field is now $\vU_0 + \vU$. Therefore, the Hall-MHD equations given in Eqs. (\ref{eq:HallMHD})-(\ref{eq:NS}) become
\begin{eqnarray}
\frac{\partial\vB}{\partial t} + u_0(x)\frac{\partial\vB}{\partial y} - \frac{d u_0}{d x} B_x \vvy & = & \bnabla\btimes\left[\left( \vU - \epsilon v_A\bnabla\btimes\vB\right)\btimes\vB\right] + \eta \nabla^2 \vB \label{eq:HallMHD-shear}\\
\frac{\partial\vU}{\partial t} + u_0(x)\frac{\partial\vU}{\partial y} + \frac{d u_0}{d x} U_x \vvy & = & - \left( \vU \cdot \bnabla \right) \vU + v_A^2 \left( \bnabla\btimes\vB\right) \btimes \vB - \bnabla P  + \nu \nabla^2 \vU + \vF . 
\label{eq:NS-shear}
\end{eqnarray}
Often times such a shear flow is meant to simulate a large scale velocity gradient acting on the relatively more microscopic degrees of freedom of the flow dynamics. We assume an imposed large-scale flow given by
\begin{equation}
u_0(x) = U_0 \left[ \tanh\left( \frac{x-\frac{\pi}{2}}{\Delta} \right) - \tanh \left(\frac{x-\frac{3\pi}{2}}{\Delta} \right) - 1 \right] ,
 \label{eq:tanh-shear}
\end{equation}

\begin{figure}
\includegraphics[width=11cm]{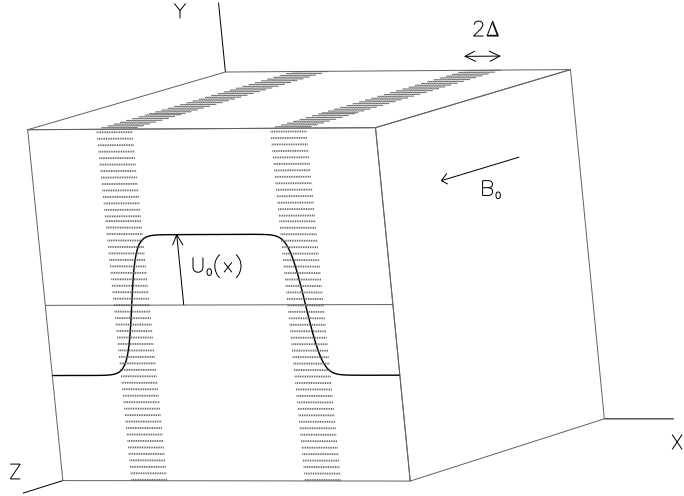}% Here is how to import EPS art
\caption{\label{fig:fig_1} Numerical box displaying the imposed velocity flow $U_y(x)$ and the external magnetic field $B_0 \vvz$. The shaded patches correspond to regions with intense shear. Each axis ranges from $0$ to $2\pi$.}
\end{figure}
which corresponds to the encounter of largely uniform flows of intensities $+ U_0 \vvy$ and $- U_0 \vvy$ through an interface of thickness $2\Delta $ parallel to the flows. The configuration is sketched in Figure~\ref{fig:fig_1}, where the jump provided by the
hyperbolic tangent is duplicated to satisfy periodic boundary conditions throughout the numerical box. 

The assumption of a hyperbolic tangent profile for shear flows with a finite thickness is standard practice in the literature \cite{drazin_1958,chandra1961,miura_1992}, as a way to study the evolution of such flows in a simplified configuration. The velocity profile given in Eqn.~(\ref{eq:tanh-shear}) is an exact equilibrium of Eqs.~(\ref{eq:HallMHD})-(\ref{eq:NS}) obtained through the application of the stationary external force $\vF_0 = -\nu\nabla^2 u_0(x)\vvy$, in the absence of magnetic field (or more generally, in the presence of a uniform magnetic field along $\vvz$).  Since the initial profile would slowly diffuse because of the effect of the viscous force, it will not be an exact equilibrium of the equation of motion. Our way out of this technical difficulty is therefore to apply a stationary force that reaches an equilibrium with the viscous force. In equilibrium, the work exerted by this force on the fluid exactly compensates for the viscous energy dissipation. In the ideal limit, this stationary force will become asymptotically zero. This strategy provides a reasonable numerical description of large scale astrophysical flows for which the effect of viscosity is negligibly small. In practice, it amounts to situations such that the relevant timescales are much shorter than the diffusion time for the large scale flow. We therefore apply this external force in our numerical box, to make sure that we are numerically studying the stability of the equilibrium given by the velocity profile of Eqn.~(\ref{eq:tanh-shear}).

For sufficiently small parcels of fluid near the center of the shaded regions displayed in Figure~\ref{fig:fig_1}, the external shear can be approximated by a linear profile given by,
\begin{equation}
\vU_0 \approx \frac{U_0}{\Delta} (x-x_0) \vvy = \omega_{sh} (x-x_0) \vvy\ ,
\label{eq:linear-shear}
\end{equation}
which corresponds to a flow of constant vorticity of intensity $\omega_{sh}$ pointing in the $\vvz$ direction in the slice centered 
at $x_0=\pi/2$ (and constant vorticity $- \omega_{sh} \vvz$ in the slice centered at $x_0=3\pi/2$). The dynamics of plasmas embedded in linear shear profiles has been numerically studied using the so-called \textit{shearing-box} simulations \cite{Hawley1995,Branden2002}. For the particular case of Hall-MHD flows, one-dimensional \textit{shearing-box} simulations have also been reported to study shear-driven instabilities \cite{bejarano2011}.  

In what follows, we also assume the plasma to be immersed in a uniform magnetic field given by $B_0 \vvz$, so that the total magnetic field is given by $B_0 \vvz + \vB$. As mentioned, the equilibrium velocity profile given by Eqn.~(\ref{eq:tanh-shear}) is an exact solution of Eqs.~(\ref{eq:HallMHD})-(\ref{eq:NS}) even in the presence of a uniform magnetic field $B_0 \vvz$ and under the action of the 
external force $\vF_0 = -\nu\nabla^2u_0(x)\vvy$. Therefore, the vector fields $\vU$ and $\vB$ hereafter correspond to the departures from this exact equilibrium.

\section{\label{sec:linear}Linearised Hall-MHD equations}

The linearised version of Eqs.~(\ref{eq:HallMHD-shear})-(\ref{eq:NS-shear}) to describe the dynamics of the perturbative components $\vU$ and $\vB$ are

\begin{eqnarray}
\frac{\partial\vB}{\partial t} + u_0\frac{\partial\vB}{\partial y} - u'_0 B_x \vvy & = & \bnabla\btimes\left[\left( \vU - \epsilon v_A\bnabla\btimes\vB\right)\btimes\vvz\right] + \eta \nabla^2 \vB \label{eq:HallMHD-linear}\\
\frac{\partial\vU}{\partial t} + u_0\frac{\partial\vU}{\partial y} + u'_0 U_x \vvy & = & v_A^2 \left( \bnabla\btimes\vB\right) \btimes\vvz - \bnabla P  + \nu \nabla^2 \vU , 
\label{eq:NS-linear}
\end{eqnarray}
where $u'_0$ expresses the spatial derivative of the profile $u_0(x)$.

This linear set of equations contains two competing instabilities: the hydrodynamic and large-scale \textit{Kelvin-Helmholtz} instability, and the magnetohydrodynamic and small-scale \textit{Hall-MSI} instability. Both of them are shear-driven instabilities, i.e., they arise 
in the shaded regions shown in Figure~\ref{fig:fig_1}. In the next two sections we summarize the basic features of 
each of these instabilities.

\section{\label{sec:kh} Kelvin-Helmholtz instability (KHI)}
A shear flow such as the one given by Eqn.~(\ref{eq:tanh-shear}) is subjected to the well known Kelvin-Helmholtz instability (KHI), 
which is of a purely hydrodynamic nature, i.e., it occurs even in the absence of any magnetic field. Within the framework of MHD, the 
stability of a tangential velocity discontinuity (i.e. in the limit of $\Delta =0$) was first studied by Ref.~[\onlinecite{chandra1961}]. For the case of an external magnetic field aligned with the shear flow, the mode is stabilized by the magnetic 
field, unless the velocity jump exceeds twice the Alfv\'en speed. For the case we are interested in, i.e., an external magnetic 
field perpendicular to the shear flow (see Figure~\ref{fig:fig_1}), the magnetic field has no effect and the flow is unstable for 
all velocity jump intensities. 

A stability analysis of a sheared MHD flow of finite thickness (i.e., 
$\Delta \ne 0$) in a compressible plasma has also been performed \cite{miura1982}, confirming the result that an external magnetic field perpendicular to the shear flow 
has no effect on the KHI, i.e., it reduces to the hydrodynamic case. If we approximate the hyperbolic tangent profile given 
in Eqn.~(\ref{eq:tanh-shear}) by piecewise linear functions, the instability growth rate arising from Eqn.~(\ref{eq:NS-linear}) is 
(see Ref.~[\onlinecite{drazin1981}])
\begin{equation}
\gamma_{kh}^2 = \frac{1}{4\Delta^2}\left(e^{-4k_y\Delta} - (2k_y\Delta - 1)^2\right) ,
\label{eqn:gamma-kh}
\end{equation}
which attains its maximum at $\lambda_{max}\approx 15.7\ \Delta$ and $\gamma_{kh,max}\approx 0.2/\Delta$.
\begin{figure}
\hfill\includegraphics[width=6.5cm]{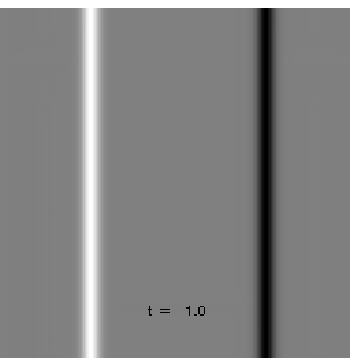}\hfill\includegraphics[width=6.5cm]{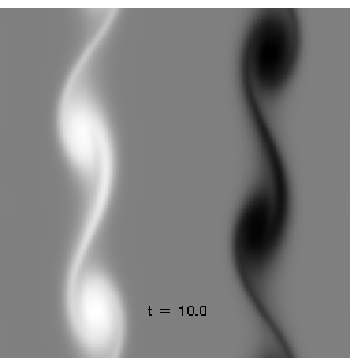}\hfill% Here is how to import EPS art
\caption{\label{fig:fig_2} Distribution of $\omega_z$ on the $(x,y)$ plane for the MHD run ($x$ is the horizontal axis and each axis ranges from $0$ to $2\pi$) at an early time $t=1$ (left frame) and also 
at a later time $t=10$ (right frame), where the Kelvin-Helmholtz instability has entered a nonlinear stage. Light (dark) regions correspond to structures of strong positive (negative) vorticity $\omega_z$.}
\end{figure}
In our dimensionless units, the numerical box has linear size $2\pi$. We set $\Delta = 0.1$ so that the instability rate peaks 
at $\lambda_{max} \approx \pi/2$. 

\begin{table}[b]%The best place to locate the table environment is directly after its first reference in text
\caption{\label{tab:table1}%
Values of dimensionless parameters for runs MHD and HMHD: $v_A$ is the Alfv\'en speed, $\Delta$ is the initial thickness of the shear layer, $\eta$ is the magnetic diffusivity, $\nu$ is the kinematic viscosity, and $\epsilon$ is the the ion skin depth.
}
\begin{ruledtabular}
\begin{tabular}{lccccc}
\textrm{Run}& $v_A$ & $\Delta$ & $\eta$ & $\nu$ & $\epsilon$ \\
\colrule
MHD & 1 & 0.1 & $2.10^{-3}$ & $2.10^{-3}$ & 0.0 \\
HMHD & 1 & 0.1 & $2.10^{-3}$ & $2.10^{-3}$ & 0.4 \\
\end{tabular}
\end{ruledtabular}
\end{table}

We perform a numerical integration of Eqs.~(\ref{eq:HallMHD-shear})-(\ref{eq:NS-shear}) subjected to the shear profile given in Eqn.~(\ref{eq:tanh-shear}) on the cubic box of linear size $2\pi$ sketched in Figure~\ref{fig:fig_1}, assuming periodic boundary conditions in all three directions. The number of gridpoints is $256^3$ and the dimensionless Alfven speed was set at $v_A = 1$ in all our simulations, indicating that the external magnetic field intensity $B_0$ is such that its Alfven velocity is comparable to the maximum velocity $U_0$ of the shear profile. The values of the dimensionless parameters required for these simulations are listed in Table~\ref{tab:table1}, both for purely MHD simulations (i.e. $\epsilon = 0$) and for HMHD simulations.  We use a pseudospectral strategy to perform the spatial derivatives and a second order Runge-Kutta scheme for the time integration (see a detailed description of the code in \cite{gomezetal2005}). For the viscosity and resistivity coefficients we chose $\nu = \eta = 2.10^{-3}$ (see Table~\ref{tab:table1}), which are small enough to produce energy dissipation only at very small scales, comparable to the Nyquist wavenumber. In all simulations, the pressure in Eqn.~(\ref{eq:NS}) is obtained self-consistently by taking the divergence of the equation, using the incompressibility condition, and solving at each time step the resulting Poisson equation for the pressure.

\begin{figure}
\includegraphics[width=11cm]{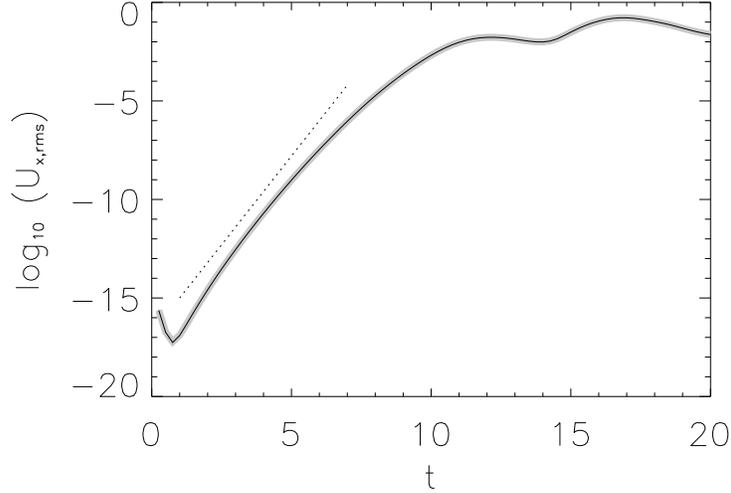}% Here is how to import EPS art
\caption{\label{fig:fig_3} R.M.S. value of $<U_x^2>_{y,z} (x_0,t)$ vs. time in a lin-log plot for the MHD run (i.e. $\epsilon = 0$). The thin black trace corresponds to $x_0=\pi/2$ and the thick gray trace to $x_0=3\pi/2$. Superimposed we show a fit corresponding to $\gamma_{kh} \approx 1.8$.}
\end{figure}

The evolution of the $\vvz$-component of vorticity is shown in Figure~\ref{fig:fig_2} at two different times for an MHD run ($\epsilon = 0$, see Table~\ref{tab:table1}). To estimate the instability growth rate, we use the component $U_x$ evaluated at $x_0=\pi/2, 3\pi/2$ (i.e., in the central part of the shear flows) as a proxy. In Figure~\ref{fig:fig_3} we show the r.m.s. value of $<U_x^2>_{y,z} (x_0,t)$ (i.e., averaging over all values on the $(y,z)$-plane) vs. time, 
\begin{equation}
U_{x,rms}^2(x_0,t) = <U_x^2>_{y,z}(x_0,t) = \int_0^{2\pi} dy \int_0^{2\pi} dz\ U_x^2(x_0,y,z,t)  .
 \label{eq:urms}
\end{equation}
The thin black trace corresponds to $x_0 = \pi/2$, while the thick gray trace corresponds to $x_0 =3\pi/2$, although (as expected) the two curves are almost undistinguishable.

Our best fit to this exponential growth, corresponds to $\gamma_{kh} \approx 1.8$, which is fully consistent with the theoretical value given 
in Eqn.~(\ref{eqn:gamma-kh}), considering that the spatial spectral content of $U_x (x_0=\pi/2,y,z)$ in terms of $k_y$ is peaked at $k_y=2,3$.

\section{\label{sec:hmsi} Hall Magneto-shear instability (Hall-MSI)}

In its simplest version, the Hall-MSI instability arises on $(x,z)$-planes of the configuration depicted in Figure~\ref{fig:fig_1}, i.e., assuming translational symmetry along the $\vvy$-direction ($\partial_y = 0$). For this instability 
to occur, we need the Hall term to be non-negligible ($\epsilon \ne 0$ in Eqn.~(\ref{eq:HallMHD-linear})) and also an intense shear. Therefore, this instability will be spatially localized around $x_0=\pi/2$ and $x_0=3\pi/2$ (i.e., the shaded slices shown in Fig.~\ref{fig:fig_1}).
Within these slices, we approximate the imposed shear flow by a linear profile characterized by a constant external vorticity 
$\omega_{sh}~\vvz$, as shown in Eqn.~(\ref{eq:linear-shear}). Note that $\omega_{sh} = +U_0/\Delta$ at $x_0=\pi/2$, corresponding to 
a vorticity vector field aligned with the external magnetic field $B_0~\vvz$. On the other hand $\omega_{sh} = -U_0/\Delta$ 
at $x_0=3\pi/2$, which implies that vorticity is anti-parallel to the magnetic field in this slice. Under these considerations, the linear equations~(\ref{eq:HallMHD-linear})-(\ref{eq:NS-linear}) lead to the following dispersion relation: 
\begin{equation}
\left(\frac{\gamma}{v_A k_z}\right)^4 + \left(\frac{\gamma}{v_A k_z}\right)^2 \left(2 + \frac{\omega_{sh}\epsilon}{v_A} + \epsilon^2 k^2\right) + \left(1 + \frac{\omega_{sh}\epsilon}{v_A}\right) = 0 ,
 \label{eq:gamma_hmsi}
\end{equation}
where $k_z \ne 0$ and $k^2 = k_x^2 + k_y^2 + k_z^2$ and $\omega = i\gamma$.  We note that the dispersion relationship displayed in Eqn.~(\ref{eq:gamma_hmsi}) has also been obtained (see Eqn.~(38) in Kunz 2008) for weakly ionized plasmas embedded in shear flows in the limit of asymptotically large density of neutrals, for which ambipolar diffusion becomes negligible in comparison with the Hall term \cite{kunz2008}. The solutions of Eqn.~(\ref{eq:gamma_hmsi}) are
\begin{equation}
\left(\frac{\gamma}{v_A k_z}\right)^2 = -\left(1 + \frac{\omega_{sh}\epsilon}{2 v_A} + \frac{1}{2}\epsilon^2 k^2\right)\ \pm\sqrt{\left(1 + \frac{\omega_{sh}\epsilon}{2 v_A} + \frac{1}{2}\epsilon^2 k^2\right)^2-\left(1 + \frac{\omega_{sh}\epsilon}{v_A}\right)}
 \label{eq:disprel}
\end{equation}

\begin{figure}
\includegraphics[width=11cm]{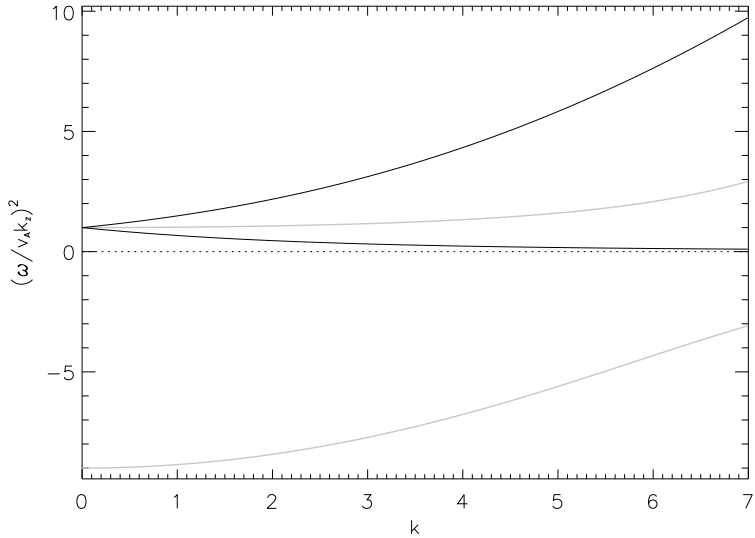}% Here is how to import EPS art
\caption{\label{fig:fig_4} Black lines show $(\omega/v_Ak_z)^2$ versus $k$ for the normal modes in the absence of shear (i.e. $\omega_{sh}= 0$). The upper branch corresponds to {\it whistlers}, while the lower one corresponds to the {\it ion-cyclotron} mode. The gray lines show the modified solutions in the presence of shear, more specifically for $\omega_{sh}=-10$. The negative branch corresponds to unstable evolutions, since $\omega^2 < 0$.
}
\end{figure}

In the absence of shear (i.e., $\omega_{sh} = 0$), this dispersion relation does not correspond to any instability, since its solutions satisfy $\gamma^2 < 0$. The solutions for $\omega_{sh} = 0$ are shown by the two black lines in Figure~\ref{fig:fig_4} and in fact the positive branch in Eqn.~(\ref{eq:disprel}) corresponds to the propagation of \textit{whistlers} (right-hand circularly polarized, upper black line) while the negative branch corresponds to \textit{ion-cyclotron} waves (left-hand polarized, lower black line), which are the normal modes for incompressible Hall-MHD \cite{gomezetal2008}. These modes propagate in any arbitrary direction, except those exactly perpendicular to the external magnetic field (since $k_z \ne 0$). 

The necessary and sufficient condition for instability is that the last term in Eqn.~(\ref{eq:gamma_hmsi}) becomes negative, namely that 
\begin{equation}
\omega_{sh} < - \frac{v_A}{\epsilon} ,
\label{eq:hmsi-cond}
\end{equation}
which renders the ion-cyclotron branch unstable. The gray curves in Figure~\ref{fig:fig_4} show the solutions of the dispersion relation 
(see Eqn.~(\ref{eq:gamma_hmsi})) modified by the presence of shear, more specifically for $\omega_{sh} = -10$. The Hall-MSI arises whenever any of the gray branches in Figure~\ref{fig:fig_4} becomes negative, since that condition would correspond to $\omega^2 < 0$ (equivalent to $\omega = \pm i\gamma$ for $\gamma > 0$). This condition can only be satisfied on the slice centered at $x_0=3\pi/2$, since its 
vorticity is negative, and not on the slice located at $x_0=\pi/2$ (since $\omega_{sh} > 0$ within that slice). Also, only the ion-cyclotron branch leads to instability, while the whistler branch remains as a propagating mode, regardless of the intensity and orientation of the shear flow. Note that although this dispersion relation was correctly obtained by Kunz 2008, 
our interpretation on the occurrence of the instability differs from the one provided in that paper. The argument exhibited by Kunz 2008 relies exclusively on the role played by the magnetic fluctuations in the linearised induction equation (see his Eqn.~(46) and below). Such an approximation would hold in the so called electron MHD regime, corresponding to negligible kinetic energy in comparison to magnetic energy, for which only the whistler mode propagates. However, we find that Hall-MSI arises as a result of the destabilization of the ion-cyclotron branch (for which the kinetic energy of the fluctuations is comparable or larger than the magnetic energy) while the whistler branch always remains stable.

\begin{figure}
\hfill\includegraphics[width=6.5cm]{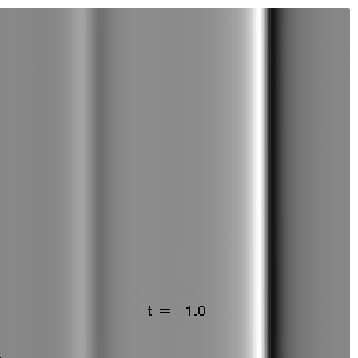}\hfill\includegraphics[width=6.5cm]{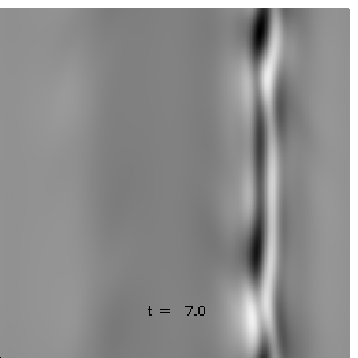}\hfill% Here is how to import EPS art
\caption{\label{fig:fig_5} Distribution of $J_z$ on the $(x,y)$ plane for the HMHD run ($x$ is the horizontal axis and each axis ranges from $0$ to $2\pi$) at an early time $t=1$ (left frame) and also 
at a later time $t=7$ (right frame). Light (dark) regions correspond to structures of strong positive (negative) electric current density $J_z$.}
\end{figure}
\begin{figure}
\hfill\includegraphics[width=6.5cm]{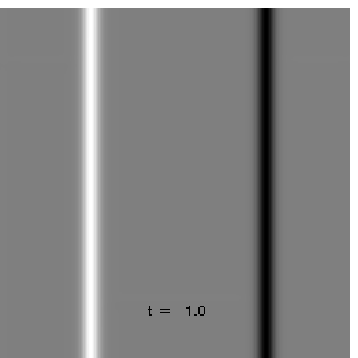}\hfill\includegraphics[width=6.5cm]{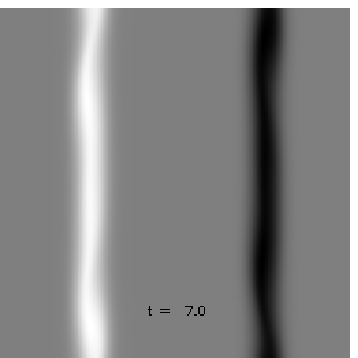}\hfill% Here is how to import EPS art
\caption{\label{fig:fig_6} Distribution of $\omega_z$ on the $(x,y)$ plane for the HMHD run ($x$ is the horizontal axis and each axis ranges from $0$ to $2\pi$) at an early time $t=1$ (left frame) and also 
at a later time $t=7$ (right frame). Light (dark) regions correspond to structures of strong positive (negative) vorticity $\omega_z$.}
\end{figure}

In Figure~\ref{fig:fig_5} we show the spatial distribution of the $\vvz$-component of the current density at two separate times, 
clearly showing the growth of the Hall-MSI on the slice containing negative vorticity centered at $x_0=3\pi/2$ for a HMHD run (i.e. $\epsilon \ne 0$, see Table~\ref{tab:table1}). Since $U_0 = 1$ and $\Delta = 0.1 $, then $\omega_{sh} = -10$ at $x_0=3\pi/2$, thus 
satisfying Eqn.~(\ref{eq:hmsi-cond}). This same simulation is also undergoing the Kelvin-Helmholtz instability, which can be 
observed in the distribution of $\omega_z (x,y)$ in Figure~\ref{fig:fig_6}.
Note that even though Kelvin-Helmholtz evolves on both slices, as shown by the vorticity patterns in Fig.~\ref{fig:fig_6}, the current 
density on the left slice (centered at $x_0=\pi/2$) remains completely unaffected, as expected. Therefore, the current density pattern formed on the right slice (see the right panel of Fig.~\ref{fig:fig_6}) is exclusively a consequence of the Hall-MSI.

According to the dispersion relation shown in Eqn.~(\ref{eq:gamma_hmsi}) and once the instability condition given by Eqn.~(\ref{eq:hmsi-cond}) is satisfied, all wavenumbers are unstable (see Figure~\ref{fig:fig_4}). The shear flow is localized, i.e., at $x_0=\pi/2$ and $x_0=3\pi/2$, where
\begin{equation} 
B_{rms}^2 (x_0,t) = <|{\vB}|^2>_{y,z}(x_0,t) = \int_0^{2\pi} dy \int_0^{2\pi} dz\ |{\vB}|^2(x_0,y,z,t) .
 \label{eq:brms}
\end{equation}

\begin{figure}
\includegraphics[width=11cm]{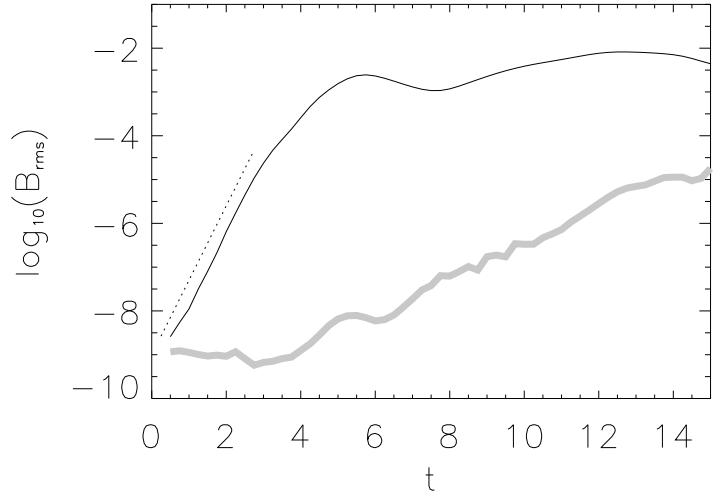}% Here is how to import EPS art
\caption{\label{fig:fig_7} R.M.S. value of the magnetic field, i.e., the square root of $ <|{\vB}|^2>_{y,z}(x_0=3\pi/2,t)$ vs. time in a lin-log plot for the HMHD run ($\epsilon = 0.4$). Superimposed we show a fit corresponding to $\gamma_{hmsi} \approx 1.7$. In grey trace we also show the square root of $ <|{\vB}|^2>_{y,z}(x_0=\pi/2,t)$.}
\end{figure}

Figure~\ref{fig:fig_7} shows $B_{rms}(x_0=3\pi/2,t)$ as a function of time in a lin-log plot. We can observe the linear stage of the 
instability, where this function grows exponentially fast. The best fit corresponds to $\gamma_{hmsi} \approx 1.7$, which is also shown 
with a dotted trace. We also overlay $B_{rms}(x_0=\pi/2,t)$ using a grey trace, which does not reflect the Hall-MSI instability, 
since the slice centered at $x_0=\pi/2$ does not satisfy the instability condition (i.e., the inequality shown in Eqn.~(\ref{eq:hmsi-cond})).

Figure~\ref{fig:fig_7} also shows how the slice centered at $x_0 = 3\pi/2$ (black thin trace) gradually departs from the linear regime and the instability saturates giving rise to a stationary turbulent regime. The development of different regimes of Hall-MHD turbulence is studied elsewhere, and is beyond the scope of the present paper. For instance, \cite{mininni_2003, mininni_2005} analyze the role of the Hall term on large-scale dynamos,  \cite{gomezetal2010} perform a similar study on small-scale dynamos for different values of the magnetic Prandtl number, while \cite{gomezetal2008,martin_2010} address the anisotropic nature of Hall-MHD turbulence in plasmas embedded in strong external magnetic fields. A detailed study of the role of all nonlinear terms of the Hall-MHD equations (including the Hall term itself) on the energy cascade arising on stationary turbulent regimes, is given in \cite{mininni_2007}.

\section{\label{sec:discu} Discussion}
In Section~\ref{sec:kh} we showed that the Kelvin-Helmholtz instability arises for large-scale modes. From Eqn.(~\ref{eqn:gamma-kh}), we can readily obtain that the unstable modes satisfy $k \le 0.64/\Delta$ and also that $k = 0.4/\Delta$ is the most unstable mode, corresponding to
\begin{equation}
\gamma_{kh,max} \simeq 0.2 \omega_{sh}\ .
\label{eq:gamma_kh}
\end{equation}

On the other hand, the instability rate for Hall-MSI is obtained from the dispersion relation given in Eqn.~(\ref{eq:gamma_hmsi}). 
When the instability condition given by Eqn.~(\ref{eq:hmsi-cond}) is satisfied, all wavenumbers become unstable. The asymptotic value of the instability rate at large wavenumbers is
\begin{equation}
\gamma_{hmsi,max} \simeq \frac{v_A}{\epsilon} \sqrt{|\omega_{sh}|\frac{\epsilon}{v_A}-1}
\label{eq:gamma_hmsi_max}
\end{equation}

Since the ratio $v_A/\epsilon$ is equal to the dimensionless version of the ion-cyclotron frequency, i.e.,
\begin{equation}
\frac{v_A}{\epsilon} = \omega_{ci}\frac{L_0}{U_0}\ ,\ \ \ \omega_{ci} = \frac{eB_0}{m_ic}
\label{eq:omega_ci}
\end{equation}
\begin{figure}
\includegraphics[width=11cm]{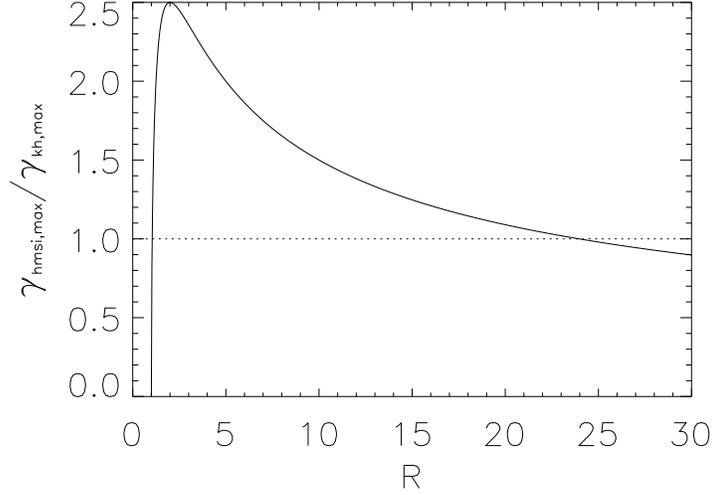}% Here is how to import EPS art
\caption{\label{fig:fig_8} Ratio of instability rates versus $R = |\omega_{sh}|/\omega_{ci}$.}
\end{figure}
the relative importance between both instability rates, depends solely on the ratio $R = |\omega_{sh}|/\omega_{ci}$. Figure~\ref{fig:fig_8} displays the ratio between the instability rates $\gamma_{hmsi,max}$ and $\gamma_{kh,max}$ as a function of $R$, showing that Hall-MSI grows faster than KH for all values of $R$ such that $1.04 < R < 23.96$. The instability condition for Hall-MSI given in  Eqn.~(\ref{eq:hmsi-cond}) is equivalent to $R > 1$, and now we find that for $R > 1.04$ it is already growing faster than KH. For $R=2$, the ratio of instability rates reaches a maximum, so that Hall-MSI is $2.5$ times more unstable than KH. Hall-MSI remains 
more unstable up to $R = 23.96$. However, note that at large values of $R$, we are moving in the direction of smaller spatial scales, where other kinetic effects not considered in this analysis might also become relevant. 

The parameter $R$ was defined as a ratio of temporal frequencies, but at least for the case $v_A = 1$ can also be regarded as a ratio of lengthscales. Since $\omega_{sh} = U_0/\Delta $ and $\epsilon = c/(\omega_{pi}L_0)$, then $R = c/(\omega_{pi}\Delta v_A)$. For $v_A = 1$, $R$ is simply the ratio between the ion skin-depth $c/\omega_{pi}$ and the half-thickness $\Delta$ of the slice. Therefore, for Hall-MSI to occur, we need the half-thickness of the shear flow to be somewhat thinner than the ion skin-depth. The thickness of shear flows will depend on the particular problem, and there are several mechanisms in astrophysics and space physics that may generate shear at very small scales and therefore might drive Hall-MSI. For instance, a whole range of sizes will spontaneously develop in turbulent flows down to the dissipative structures where viscosity becomes dominant. This is the case in the solar wind, where fluctuations are observed well below the 
ion skin depth. Besides turbulence, other examples of thin shear flows are differential rotation in accretion disks \cite{balbus_2001}, zonal flows in drift wave turbulence in tokamaks \cite{pushkarev_2013} or magnetic reconnection in shear flows \cite{knoll_2002}.

Note that the ratio of instability rates depicted in Figure~\ref{fig:fig_8} corresponds to the idealized case where the KHI is dominated by its most unstable mode and Hall-MSI is led by very small wavelength modes. In general, both $\gamma_{kh}$ and $\gamma_{hmsi}$ will depend explicitly on the wavenumber, and therefore the ratio $R$ will be determined also by the initial condition. For instance, for the simulations shown in this work (corresponding to $R\approx 4$), both instability rates turned out to be comparable, while according to Figure~\ref{fig:fig_8}, the ratio would be different from unity. 

\section{\label{sec:conclu}Conclusions}

In the present paper, we performed a comparative study of two competing shear-driven instabilities in a fully ionized plasma: Kelvin-Helmholtz and the Hall magneto-shear instability. Kelvin-Helmholtz is probably the paradigm of shear-driven instabilities, which leads to a large scale corrugation of the shear layer, regardless of the presence of perpendicular magnetic fields. This instability has been invoked to play a role in several astrophysical plasmas, such as near the boundaries between astrophysical jets and the interstellar surroundings. On the other hand, in sufficiently low density plasmas, also the Hall magneto-shear instability can take place, in which the Hall effect and the presence of a magnetic field play essential roles. Therefore, we carried out three-dimensional simulations of the Hall-MHD equations, starting from configurations such that these two instabilities are present. 

The main result reported in this paper is that when the shear flow is intense enough that its central vorticity surpasses the ion-cyclotron frequency of the plasma, the Hall magneto-shear instability becomes non-negligible. Furthermore, we show that Hall-MSI has growth rates larger than those for KHI for a wide range of values of the parameter $R$, which is the ratio between the vorticity at the center of the shear layer and the ion-cyclotron frequency of the plasma. We therefore believe that this result might have an impact on several astrophysical shear flows, such as the above mentioned example of astrophysical jets. This unexpected result is a direct consequence of the existence of a relatively new instability (namely, Hall-MSI, see \cite{kunz2008}, also \cite{bejarano2011}), which shows the potencial relevance of the Hall effect in highly sheared plasma flows.

\bibliography{gomez}% Produces the bibliography via BibTeX.

\end{document}